\documentclass[12pt]{article}

\def\be{\begin{equation}}
\def\ee{\end{equation}}
\def\bea{\begin{eqnarray}}
\def\eea{\end{eqnarray}}

\def\re#1{(\ref{#1})}

\begin{document}
\begin{titlepage}

\vskip 1. true cm
\begin{center}
\large{
{Fock Space Representations for Non-Hermitian Hamiltonians}}
{ \vskip 0.6 true cm
{ David B. Fairlie$^1$\ \ and\ \  Jean Nuyts$^2$}
\vskip 0.5 true cm
{\small
\centerline{David.Fairlie@durham.ac.uk, Jean.Nuyts@umh.ac.be}}
\vskip 0.2 true cm
{\it $^1$ Department of Mathematical Sciences, University of Durham}\\
{\it South Road, DH1 3LE, Durham, England}\\[3pt]
{\it $^2$  Physique Th\'eorique et Math\'ematique,
Universit\'e de Mons-Hainaut}\\
{\it 20 Place du Parc, 7000 Mons, Belgium}}
\vskip 1.3 true cm
\noindent
{\bf Abstract}
{\small\begin{flushleft} The requirement of Hermiticity of a  Quantum Mechanical Hamiltonian, for the
description of physical processes with real eigenvalues which has been challenged
notably by Carl Bender, is examined for the case of a Fock space
Hamilitonian which is  bilinear in two creation and destruction operators. An
interpretation of this model as a Schr\"odinger operator leads  to an
identification of the Hermitian form of the Hamiltonian as the Landau model of a
charged particle in a plane, interacting with a constant magnetic field at right
angles to the plane. When the parameters of the Hamiltonian are suitably
adjusted to make it non-Hermitian, the model represents two harmonic oscillators
at right angles interacting with a constant magnetic field in the third
direction, but with a pure imaginary coupling, and real energy eigenvalues. It
is now ${\cal PT}$ symmetric. Multiparticle states are investigated.
\end{flushleft}}
\end{center}
\vfill
\end{titlepage}

\section{Introduction \label{secIntrod}}
The familiar Hamiltonians of Quantum Mechanics may be analysed for symmetries
either in terms of pure matrix algebra, or else in terms of a Fock space
representation, which generally leads to a more physical interpretation of the
mathematical manipulations. In particular, any  Hamiltonian constructed from a
Fock space of $n$   fermionic creation and annihilation operators may be
transcribed in terms of a finite dimensional matrix of dimension
$2^\frac{n}{2}\times 2^\frac{n}{2}$ on account of the fact that such operators
can always be constructed from gamma matrices, which admit a well known matrix
representation.
The inverse of this construction,  to obtain a Fock space representation of any
finite dimensional Hermitian matrix can be done by embedding the matrix in a
larger one of suitable dimension to admit representations of the canonical
anti-commutation relations.
In fact, in view of the existence of a matrix representation for Hamiltonians
the transcription is merely an exercise in matrix algebra for undergraduates.
However, the situation is more complicated when the Hamiltonian is no longer
Hermitian. It is a central tenet of Quantum Theory and Quantum Field Theory that
the Hamiltonian should be Hermitian, but this is not in fact necessary to
maintain real eigenvalues, as has been demonstrated in recent years, notably by
Bender and Boettcher.\cite{bender}.
 However, if the eigenvalues are all discrete and real and their eigenvectors
span the full space, there will exist, in general, an infinity of Hermitian
Hamiltonians (possibly infinite dimensional) which are unitarily equivalent
to the diagonal (real) form. They are equivalent up to a more general (not
necessarily unitary) transformation to the initial Hamiltonian.

 If all the eigenvalues are real and their eigenvectors span a subspace $S$
of the initial space (say finite dimensional to make things simpler) there
exists, in general, an infinity of Hamiltonians unitarily equivalent to a
Jordan form (or a generalised Jordan form in the infinite dimensional case)
and equivalent up to a more general (non necessary unitary) to the initial
Hamiltonian. These Hamiltonians projected to the subspace $S$ are Hermitian.

If some of the eigenvalues are discrete and real, this same statement will
obviously apply in the subspace of the corresponding eigenvectors.
 We shall study a simple example which illustrates this in a physical context,
that of the Landau problem of a particle in two dimensions moving under the
infuence of a constant magnetic field in the third direction\cite{gamboa}\cite{
dayi}, and show that this model exhibits  the features which have been found for
${\cal PT}$ symmetric non-Hermitian systems. In particular, while the usual
model is Hermitian, with a real magnetic coupling,
a deformation of the parameters in this model leads under certain systems to a
non-Hermitian system with real eigenvalues and an imaginary coupling. In this
respect it recalls to mind the work of Hollowood \cite{hollowood}, who showed
that  affine Toda field theory with pure imaginary coupling possesses real
energy levels.
This system represents an anisotropic oscillator in interaction with a constant
magnetic field. Some discussion is given of the case of transition between the
Hermitian and non-Hermitian case, where two of the eigenvalues coincide,
and the associated matrix cannot be fully diagonalised, but can be reduced by
similarity transformations only to Jordan Normal form.

\section{Two-dimensional Case \label{secboson}}
Take a $2\times 2$ matrix
\be
M\,=\,\left[ \begin{array}{cc}
                s_1 & s_3\\
                s_4 & s_2
\end{array}\right]\ .
\label{mat}
\ee
Then  any finite dimensional matrix eigenvalue problem can always be transformed
to an infinite dimensional
Quantum Mechanical eigenvalue problem by the introduction of Fock space creation
and annihilation operators
$a^\dag_i,\ a_j$,  so that matrix $M_{ij}$ turns into the Hamiltonian
$\sum_{i,j}a^\dag_iM_{ij}a_j$. This gives in this case the Hamiltonian
\be
H=s_1a^{\dag}_1a_1+s_2a^{\dag}_2a_2+s_3a^{\dag}_1a_2+s_4a^{\dag}_2a_1\ ,
\label{ham}
\ee
where the coefficients $s_i,\ i=1\dots 4$ are complex constants and
$a^{\dag}_i,\ a_i$ are two creation and annihilation operators satisfying the
usual canonical commutation relations, with  only the following non zero
commutators,
\be
a_ia^{\dag}_j-a^{\dag}_ja_i=\delta_{ij}\ .
\label{com}
\ee
The vacuum state $\mid 0>$ is defined as usual as the normalized
state which is annihilated by the $a$'s
\be
a_i\mid 0> = {\cal{O}}
\label{vacuum}
\ee
where ${\cal{O}}$ is the zero norm state in the Hilbert space.

\section{Diagonalisation \label{secdiag}}
Define
\be
\Delta=\sqrt{(s_1-s_2)^2+ 4s_3s_4};\ \ \ \lambda_{\pm}=\frac{1}{2}(s_2-s_1)\pm
\frac{1}{2}\Delta \ ,
\label{def3}
\ee
and construct the linear combinations
\bea
\alpha_1^{+} &=& n_1\left(\frac{s_3}{\lambda_{+}}a_1^\dag+a_2^\dag\right)\label
{def4}\\
\alpha_2^{+} &=& n_2\left(\frac{s_3}{\lambda_{-}}a_1^\dag+a_2\dag\right)
\label{def5}\\
\alpha_1 &=&\frac{1}{n_1}\left( \frac{s_4}{\Delta}a_1+\frac{\lambda_{+}}{\Delta}
a_2\right)
\label{def6}\\
\alpha_2 &=& \frac{1}{n_2}\left(\frac{-s_4}{\Delta}a_1+\frac{-\lambda_{-}}{
\Delta}a_2\right)\ .
\label{def7}
\eea
In this construction, $\alpha_j^{+}$ denotes a creation operator, which should
not be thought of as the Hermitian conjugate of $\alpha_j$
which is also a destruction operator, since there is no Hermitian
Hamiltonian ($\alpha_j^{\dag}\neq \alpha_j^{+}$).
However the properties
\be <0\mid \alpha_i\alpha_j^{+}\mid 0> \,=\,\delta_{ij}
\label{orth}
\ee
hold.  Indeed the $\alpha_j$'s and $\alpha_j^{+}$'s satisfy exactly the same
commutation relations as do the $a$'s, as a consequence of the fact that the
matrix (\ref{mat}) can be diagonalised by a similarity transform.

 This implies that observables in the theory can be calculated, as $n$ particle
 states will have the form
\be
\mid r,n-r>\,=\,\frac{1}{\sqrt{r!(n-r)!}}\left(\alpha_1^{+}\right)^r\left(
\alpha_2^{+}\right)^{n-r}\mid 0>\ .
\label{orth2}
\ee
These states are orthogonal to the following adjoint states
\be
\mid{\rm adj}[r, n-r]> \,=\,
\frac{1}{\sqrt{r!(n-r)!}}
\left(\alpha_1^{\dag}\right)^r
\left(\alpha_2^{\dag}\right)^{n-r}
\mid 0>
\label{orth3}
\ee
in the sense that
\be
<{\rm adj}[p', q']\mid p,q> \,=\, \delta_{p'p}\delta_{q'q} \ .
\label{orth5}
\ee
The limitation of the use of non-Hermitian Hamiltonians is that
the notion of $+$ conjugation and reality of eigenvalues is specific to the
Hamiltonian used and is not
universal as is the case with Hermitian conjugation,  as Bender et al have
remarked \cite{bender0}. Hermiticity also guarantees reality of eigenvalues,
independently of details of the Hamiltonian. This universality is a consequence
of the fact that the
inverse of a unitary matrix, which diagonalises a Hermitian matrix, is its own
Hermitian conjugate, i.e. since $U^\dag \,=\,U^{-1 }$ for a unitary matrix $U$
the columns of $U^\dag$ are orthogonal
to the rows of $U$,  the operation of Hermitian conjugation works independently
of the Hermitian matrix to be diagonalised.
In the case of a $2\times 2$ matrix $H$  we have
\bea
H&=&s_1a^{\dag}_1a_1+s_2a^{\dag}_2a_2+s_3a^{\dag}_1a_2+s_4a^{\dag}_2a_1
\nonumber
\\
&=&\,\frac{1}{2}(s_1+s_2 +\Delta)\alpha^{+}_1\alpha_1
+\frac{1}{2}(s_1+s_2-\Delta)\alpha^{+}_2\alpha_2\ .
\label{diaghamil}
\eea
This construction shows that the $n$ particle states have energies of the form
\be
E_{n,m}=\frac{n}{2}(s_1+s_2) +\frac{m}{2}\Delta\ ,
\label{eigens}
\ee
where $m$ runs from $-n$ to $n$ in steps of 2.  There are several interesting
features of this result. The eigenvalues will be real and distinct provided
$s_1+s_2$ is real and $(s_1-s_2)^2+ 4s_3s_4$ is positive, thus even when the
Hamiltonian is non-Hermitian, the eigenvalues may be real. When this latter
factor is zero, then
the eigenvalues are degenerate. Then,
either the Hamiltonian is proportional to the unit matrix ($s_1=s_2$,
$s_3=s_4=0$) or, if at least one off-diagonal element is non zero, there is one
eigenvector only and the Hamiltonian can be brought to the normal form
($s'_1=s'_2$, $s'_3=0$, $s'_4=1$).
These cases are briefly discussed later. For non-Hermitian Hamiltonians,
Bender et al \cite{bender0} have analysed the existence of a Parity operator
${\cal P}$ and a Conjugation matrix ${\cal PT}$.
These crucial issues will be taken up and generalised after we demonstrate a
simple application
to a physical example of the $2\times 2$ situation detailed above.

\section{Schr\"odinger interpretation \label{secSchroding}}

The above problem is equivalent up to a constant energy shift to solving a
Schr\"odinger equation for a particular Quantum Mechanical problem;
set $a_i = p_i +iq_i$ etc; then the problem is equivalent to
\bea
H &\,=\,&
    -s_1(\frac{\partial^2}{\partial x^2}-x^2)
    -s_2(\frac{\partial^2}{\partial y^2}-y^2)
		\nonumber\\
  &&-(s_3+s_4)(\frac{\partial^2}{\partial x\partial y}-xy)
    +(s_3-s_4)(x\frac{\partial}{\partial y}-y\frac{\partial}{\partial x})
\label{schrod}
\eea
up to a $c$-number addition.
If $s_3 =-s_4$,  then this is the Landau problem of a particle in a plane
coupled to a magnetic field in the perpendicular direction, through its angular
momentum with possibly an additional linear central force.  To see this consider
a constant magnetic field with potential
$\displaystyle{ \vec A\,=\, \frac{B}{2}(\ x)}$ in the Hamiltonian
\bea
H&\,=\,& \frac{1}{2m}(\vec p+\frac{e}{c}\vec A)^2
         \nonumber\\
 &\,=\,& \frac{1}{2m}\left(p_x^2+p_y^2
+\left(\frac{eB}{2c}\right)^2(x^2+y^2)+\frac{eB}{c}(p_yx-p_xy)\right)\ .
\label{ham2}
\eea
and make the identifications; $s_1=s_2=1/(2m)$, and $s_3 =-s_4$ is pure
imaginary. This Hamiltonian is Hermitian with real eigenvalues. However, the
choice of $s_3=-s_4$  and real, eliminates the $xy$ cross terms and also gives
 real eigenvalues, provided that the factor $(s_1-s_2)^2- 4s_4^2$ is positive
 which necessarily entails  that $s_1\neq s_2$.
In this  case  the Hamiltonian is non-Hermitian and it represents a constant
magnetic field coupled to an anisotropic oscillator with pure imaginary
coupling. In all the examples Bender et al \cite{bender}\cite{bender2} have
constructed with non-Hermitian Hamiltonians and real eigenvalues the Hamiltonian
is ${\cal PT}$ symmetric.   Dorey, Dunning and Tateo \cite{dorey} have recently
given a proof that the spectra of a number of ${\cal PT}$ invariant Hamiltonians
are entirely real. However, on its own  ${\cal PT}$ invariance only implies real
or complex conjugate pairs of eigenvalues. This is the situation   under
consideration here  as ${\cal PT}$ invariance means invariance under
$(x,\ y) \rightarrow (-x,\ -y)$ and
$ i\rightarrow -i$.  At the same time $\vec p$ does not change. The Hamiltonian
will then be ${\cal PT}$ symmetric since $B\rightarrow B$, but the eigenvalues
may be complex conjugate if the positivity condition is violated.

\section{Real Eigenvalue conditions\label{secreal}}
As is well known, the eigenvalues of a Hermitian matrix are always real. In
general the condition for the reality of eigenvalues depends on more specific
details of the matrix.
However in the case of an $n\times n$ matrix $H$ necessary conditions for the
existence of real eigenvalues  are given by
\be
{\rm Trace}\left(H^r\right)\,=\,{\rm real};\ \ r =1,2,\dots,n\ .
\label{herm}
\ee
This result follows simply from the observations that the characteristic
polynomial of $H$, being an invariant under similarity transformations, has
coefficients expressible in terms of the traces of powers of $H$  and that if
all the eigenvalues are real, $H^\dag$ must have the same eigenvalues as $H$. A
further necessary and sufficient condition for an arbitrary $2\times 2$ matrix
to possess real eigenvalues is given by the requirement that $ 2{\rm Trace}H^2 -
({\rm Trace}H)^2 \geq 0$. This expression is just a translation into invariant
form of the
condition of section (\ref{secboson}). Unfortunately, an invariant criterion
even in the 3 dimensional case is rather complicated.
When the traces are real and $\Delta=0$, the two dimensional matrix $H$ may be
expressed as
\be
 H\,=\,\frac{1}{2}(s_1+s_2)1\!\!1+A\label{jordan1}
\ee
where $A$ is null or the zero matrix; i.e ${\rm Trace}A = {\rm Trace}A^2=0$ or
$A=0$.
In the former case the matrix is equivalent by a change of basis to a Jordan
Normal Form
\be
H\,=\,\left[ \begin{array}{cc}
                a & 1\\
                0 & a
\end{array}\right]\ .\label{jordan2}\ee

\section {``Conjugation'' and ``Parity''. Finite Dimensional Case
\label{secconjug}}

\noindent In this section, we extend the notions of ``conjugation'', and/or of
``parity'' introduced in \cite{bender} to arbitrary finite dimensional
matrices representing the Hamiltonian with real eigenvalues.
In certain cases the
arguments can be extended immediately to infinite dimensional spaces (see
Section \re{secfock} for an example).
It is also not
difficult to extend them to the case where some of the eigenvalues are equal or
are
complex or
when the starting matrix $M_d$ below has parts in a Jordan form

\vskip 0.7 cm
\noindent We first recall a few well-known facts. Let $M_d$ be a real diagonal
matrix in $p$ dimensions with, for later
simplicity, all elements different
\be
\left(M_d\right)_{ij}=\mu_j\,\delta_{ij}
        \ ,\quad \mu_j=\mu_j^*
        \ ,\quad \mu_i\neq\mu_j\quad {\rm{when\ }}i\neq j\ ,
\label{lmd}
\ee
and let $N$ be an arbitrary invertible $p\times p$ complex matrix
\noindent  Let the
scalar
product  of the column eigenvectors $\psi$ and $\phi$ be defined by the obvious
\be
<\psi\mid\phi>=\psi^{\dag}\phi
\label{lmsp}
\ee
where $A^{\dag}=(A^*)^{tp}$ is the Hermitian conjugate matrix.

\noindent For any matrix $M$ which has the eigenvalues $\mu_i$ there exists a
matrix N such that
\be
M=N\,M_d\,N^{-1}\quad\Leftrightarrow\quad M_d=N^{-1}\,M\,N\ .
\label{lmh}
\ee
The $p$ vectors $\phi_i,\,i=1,\dots,p$, with components
\be
\left(\phi_i\right)_j=\delta_{ij}\ ,
\label{lmbasic}
\ee
\noindent
are obviously eigenvectors of $M_d$ with eigenvalues $\mu_i$. They are
orthonormal for the scalar product
\be
<\phi_j\mid\phi_k>=\delta_{jk}\ .
\label{lmortho}
\ee
\noindent The vectors $\psi_i=N\phi_i$ are obviously eigenvectors of the general
$M$ with the same eigenvalue.

We now present a few results providing, in the rather general case \re{lmd},
the general solution to the question of existence and construction of a suitable
scalar product $<\phi\mid\psi>_G$, of a
``parity'' operator $P$ and of a ``conjugation'' operator $C$.

\begin{enumerate}

\item
The vectors $\psi_i$ are orthonormal for the Hermitian scalar product
\re{lmspgen} defined by
\be
<\psi\mid\phi>_{_G}
    =\psi^{\dag}\,G\,\phi\ ,\quad {\rm{with}}\ \
    G=\left(N^{-1}\right)^{\dag}\,N^{-1}\ .
\label{lmspgen}
\ee
Indeed
\bea
\delta_{jk}&=&<\phi_j\mid\phi_k>
         \nonumber\\
          &=&<N^{-1}\psi_j\mid N^{-1}\psi_k>
         \nonumber\\
         &=&\psi_j^{\dag}\left(N^{-1}\right)^{\dag}N^{-1}\psi_k
         \nonumber\\
         &=&\psi_j^{\dag}G\psi_k
         \nonumber\\
         &=&<\psi_j\mid\psi_k>_{_G}\ .
\label{lmorthpf}
\eea
Because of the form of $G$ \re{lmspgen} all the requirements of the scalar
product are met. Note however that if $N$ is unitary, then $G$ is the identity
matrix; otherwise, it depends upon the structure of $N$. This means that while
Hermitian conjugation can be defined universally for Hermitian
Hamiltonians. In the non-Hermitian case, it is a more specific matrix. Two
matrices $M$ lead to the same metric $G$ if the $N$ which defines the first one
is equal to the $N$ which defines the second multiplied by an arbitrary unitary
matrix.

\item

We denote complex by ${\cal{T}}$ the symbolic operator
which performs the $c$-number conjugation,
${\cal{T}}$
\be
{\cal{T}}A=A^*{\cal{T}} .
\label{conjug}
\ee

\noindent There exist always a matrix $P$ (generically called ``parity'' as
Bender suggested)
which has the following property
\bea
\left(P{\cal{T}}\right)\, M&=&M\, \left(P{\cal{T}}\right)
          \nonumber\\
P M^*&=&MP\ .
\label{lmparityeq}
\eea
Take any invertible matrix $K$ which commutes with $M_d$,
\be
KM_d=M_dK \ .
\label{lmcenter}
\ee

In particular any diagonal matrix
will do if the eigenvalues since are all different. If some eigenvalues
are equal, $K$ may belong to the stability group of $M_d$. Then
\be
P=NK(N^*)^{-1}\ .
\label{lmparity}
\ee
Indeed
\bea
P M^*&=&MP
       \nonumber\\
P \left(N\,M_d\,N^{-1}\right)^*&=&N\,M_d\,N^{-1}P
	\quad\quad{\rm{by\ \re{lmh}}}
       \nonumber\\
P N^*\,M_d\,(N^{-1})^*&=&
N\,M_d\,N^{-1}P
        \quad\quad{\rm{by\ \re{lmd}}}
       \nonumber\\
\left(N^{-1}\, P\, N^*\right)\,M_d&=&M_d\,\left(N^{-1}P\, N^*\right)\ .
\label{lmparitypf}
\eea
Using \re{lmparity} and \re{lmparitypf}, the statement is proved.

\item
The matrix ${\cal C}$ (conjugation) is defined as having the following property
\bea
 C\, M&=&M\, C
          \nonumber\\
C^2&=&1\ .
\label{lmparityeq2}
\eea
The general solution for $C$ is
\be
C=N K_s N^{-1}
\label{lmC}
\ee
where $K_s$ is such it commutes with $M_d$ and $K_s^2=1\!\!1$.
In the generic case,
$K_s$ is a diagonal matrix with elements $\pm$ only. Up to a trivial overall
sign, there are $2^{p-1}$ independent $C$'s.
Indeed the first equation of \re{lmparityeq} gives
\bea
C M&=&MC
       \nonumber\\
C N\,M_d\,N^{-1}&=&N\,M_d\,N^{-1}C
          \quad\quad{\rm{by\ \re{lmh}}}
       \nonumber\\
\left(N^{-1}\, C\, N\right)\,M_d&=&M_d\,\left(N^{-1}C\, N\right)
\label{lmCpf1}
\eea
which implies that $N^{-1}\, C\, N=K$ commutes with $M_d$.
The second equation implies moreover that
\be
K^2=1\!\!1\ .
\label{lmCpf2}
\ee
In the two dimensional example of section \re{secboson}, $C$ is either the
identity
matrix
or the matrix
\be
{C}\,=\,\frac{1}{\Delta}\left[ \begin {array}{cc}
                   { s_1}-{ s_2}&2\,{ s_3}
\\\noalign{\medskip}2\,{ s_4}&{ s_2}-{ s_1}\end {array} \right]
\label{c}
\ee
up to sign.
\end{enumerate}

\section{More Bosonic Creation Operators}
\label{secmoreboson}

In this section, we elaborate on two or more particle states obtained
for the simplest Hamiltonian \re{ham}. We will treat in great detail the two and
four particles states.

\subsection{Two Particle States\label{twopart}}

Let us first elaborate on the two particle states constructed from the operators
of section (\re{secboson}).
Suppose that we were to start with the products $b_1^{\dag},\,i=1,2,3$ of two
creation operators
\bea
b_1^{\dag}&=&(a_1^{\dag})^2
        \nonumber\\
b_2^{\dag}&=&a_1^{\dag}a_2^{\dag}
        \nonumber\\
b_3^{\dag}&=&(a_2^{\dag})^2
        \label{thebs}
\eea
which construct the three two-particle states $b_i^{\dag}\mid 0>,\,i=
1,2,3$
when applied
to the vacuum of the $a_i$ operators.
The corresponding three-dimensional matrix $M$ is
\be
M^{[2]}=\left( \begin{array}{ccc}
                2 s_1 & s_3        & 0       \\
                2 s_4 & s_1+s_2    & 2 s_3   \\
                0     & s_4        & 2 s_2
\end{array}\right)\ .
\label{matrixMtwo}
\ee
It satisfies (with $H$ from \re{ham})
\be
\left[H,b_i^{\dag}\right]=M^{[2]}_{ji}b_j^{\dag}
\label{Hbs}
\ee
and hence
\be
H b_i^{\dag}\mid 0>=M^{[2]}_{ji}b_j^{\dag}\mid 0>\ .
\label{Hbsvac}
\ee
The diagonalisation of $M$ provides the two particle spectrum. Using
$\Delta$ from \re{def3}, we find, as expected, three eigenstates
$\beta_i^{+}$ which fulfill the eigenvalue equation (with $H$ from
\re{ham})
\be
H\beta_i^{+}\mid 0>=\mu_i\,\beta_i^{+}\mid 0>\ .
\label{Hbesvac}
\ee
They are up to a factor, obviously from \re{secboson}
\bea
\beta_1^{+}&=& - 2 s_3^2  b_1^{\dag}
               + 2 s_3( - \Delta + s_1 - s_2)  b_2^{\dag}
\nonumber\\
&&+(\Delta (s_1 -s_2) - (s_1-s_2)^2  - 2 s_3 s_4) b_3^{\dag}
       \nonumber\\
           &\propto& (\alpha_1^{+})^2
     \nonumber\\
\beta_2^{+}&=&-s_3 b_1^{\dag}+(s_1+s_2)b_2^{\dag}+s_4 b_3^{\dag}
       \nonumber\\
           &\propto& \alpha_1^{+}\alpha_2^{+}
     \nonumber\\
\beta_3^{+}&=&- 2 s_3^2  b_1^{\dag}
               + 2 s_3(  \Delta + s_1 - s_2)  b_2^{\dag}
     \nonumber\\
&&+(-\Delta (s_1 - s_2) - (s_1-s_2)^2  - 2 s_3 s_4) b_3^{\dag}
       \nonumber\\
           &\propto& (\alpha_2^{+})^2
\label{thebes}
\eea
with eigenvalues (see \re{eigens})
\be
\begin{array}{lllllll}
\mu_1&=&s_1+s_2+\Delta&=&E_{2,2}&=&2E_{1,1}
     \nonumber\\
\mu_2&=&s_1+s_2&=&E_{2,0}&=&E_{1,1}+E_{1,-1}
     \nonumber\\
\mu_3&=&s_1+s_2-\Delta&=&E_{2,-2}&=&2E_{1,-1} \ .
\end{array}
     \label{themus}
\ee

This discussion suggests the following question: how can this spectrum
be obtained directly from the operators defining the two-particle
space? We will first show that a na\"\i ve approach has strong
shortcomings.

\begin{enumerate}

\item  A first na\"\i ve answer is as follows. Introduce symbolically
the new operators ${\widetilde{b}}_i$
which are supposed, with the $b_j^{\dag}$, to satisfy the canonical
commutation relations
\be
\left[{\widetilde{b}}_i,b_j^{\dag}\right]=\delta_{ij}
\label{canonwtild}
\ee
and define a two particle Hamiltonian
$H^{[2]}_{\rm{naive}}$ as
\be
H^{[2]}_{\rm{naive}}
=b_i^{\dag}M^{[2]}_{ij}{\widetilde{b}}_j\ .
\label{Htwopart}
\ee
Then obviously but, we insist, symbolically we obtain
\be
\left[H^{[2]}_{\rm{naive}},b_i^{\dag}\right]
=M^{[2]}_{ji}b_j^{\dag}
\label{Htcom}
\ee
and on a vacuum $\mid 0^{[2]}>$ such that
\be
{\widetilde{b}}_j\mid 0^{[2]}>=0
\label{vacuumtwo}
\ee
we obtain
\be
H^{[2]}_{\rm{naive}}b_i^{\dag}\mid 0^{[2]}>
      =M^{[2]}_{ji}b_j^{\dag}\mid 0^{[2]}> \ .
\label{Htcomvac}
\ee

It is not difficult to see that it is impossible to construct
${\widetilde{b}}_i$ satisfying \re{canonwtild} out of products of two $a_j$.
However
if the vacuum $\mid 0^{[2]}>$ is thought to be the $a$ vacuum
$\mid 0>$
\be
\mid 0^{[2]}>:=\mid 0>
\label{vacident}
\ee
and the weaker condition
\be
\left[{\widetilde{b}}_i,b_j^{\dag}\right]\mid 0>
\ =\ {\widetilde{b}}_ib_j^{\dag}\mid 0>
\ =\ \delta_{ij}\mid 0>
\label{canonwtildvac}
\ee
is imposed, the ${\widetilde{b}}_i$ can be identified to be
\be
{\widetilde{b}}_1=\frac{1}{2}b_1,
\ {\widetilde{b}}_i=b_2,
\ {\widetilde{b}}_i=\frac{1}{2}b_3 \quad{\rm{in\ the\ weak\ sense}}
\label{wtbversusb}
\ee
with, obviously, $b_1=a_1^2,\ b_2=a_1 a_2,\ b_3=a_3^2$.
The equation \re{Htcomvac} then holds.

\item In a second approach,  one tries to
construct directly an $H^{[2]}$ quadratic both in $a^{\dag}_i$ and in
$a_i$ and which satisfies the basis equation \re{Htcom}
\be
\left[H^{[2]},b_i^{\dag}\right]\mid0>=M^{[2]}_{ji}b_j^{\dag}\mid0>
\label{Htcom2}
\ee
as a result of the basic $a_i$ canonical commutation relations \re{com}.

Such a Hamiltonian does not exist in the strong sense (when the vacuum is
removed in \re{Htcom2}). In the weak sense, it
exists as
\be
H^{[2]}=b_i^{\dag}N^{[2]}_{ij}b_j\ .
\label{Htwopartt}
\ee
where the matrix $N^{[2]}$ is
\be
N^{[2]}=\left( \begin{array}{ccc}
                s_1 & s_3        & 0       \\
                s_4 & s_1+s_2    & s_3   \\
                0   & s_4        & s_2
\end{array}\right)\ .
\label{matrixNtwo}
\ee
This result is obviously equivalent to the na\"\i ve approach once the
$\widetilde{b}_i$ are expressed in terms of the $b_i$ \re{wtbversusb}.

\end{enumerate}

\subsection{Four Particle States\label{fourpart}}

Let us proceed in the same way for the four particle states which can be thought
either directly as the four particle states in the original operators (
$a_i^{\dag},\, i=1,2$)
\re{com}
\be
d^{\dag}_i=(a_1^{\dag})^{5-i}\,(a_2^{\dag})^{i-1}
     \quad,\quad i=1,\dots,5
\label{fourcreat}
\ee
or as the compound states constructed out of two of the basic two particle
states $b_i^{\dag},\,i=1,2,3$
\re{thebs}
\be
\begin{array}{ccccccccccc}
\check{d}^{\dag}_1&=&(b_1^{\dag})^{2}
     &,&
\check{d}^{\dag}_2&=&b_1^{\dag}\,b_2^{\dag}
     &,&
\check{d}^{\dag}_3&=&b_1^{\dag}\,b_3^{\dag}
     \nonumber\\
\check{d}^{\dag}_4&=&(b_2^{\dag})^{2}
     &,&
\check{d}^{\dag}_5&=&b_2^{\dag}\,b_3^{\dag}
     &,&
\check{d}^{\dag}_6&=&(b_3^{\dag})^{2}\ .
\end{array}
\label{twotwocreat}
\ee

Let us note immediately that there are only five states built with the five
$d_i^{\dag}$ while there are six states built with the
six $\check{d}^{\dag}_i$. This is due to the obvious fact that
\be
\check{d}^{\dag}_3\equiv \check{d}^{\dag}_4
\label{twotwoid}
\ee
when constructed from the $a'$s.

Let us treat in succession the $d^{\dag}_i$ case and the $\check{d}^{\dag}_i$
case.

\begin{itemize}

\item
The $5\times 5$ matrix $M^{[4]}$, analogous to \re{matrixMtwo} and defined by
\be
\left[H,d_i^{\dag}\right]=M^{[4]}_{ji}d_j^{\dag}\ ,
\label{Hbsfour}
\ee
is
\be
M^{[4]}=\left( \begin{array}{ccccc}
4 s_1   &  s_3       &        0     &    0        &      0  \\
4 s_4   &3 s_1 + s_2 &    2 s_3     &    0        &      0  \\
 0    &  3 s_4       & 2 (s_1 + s_2)&  3 s_3      &      0  \\
 0    &    0         &    2 s_4     & s_1 + 3 s_2 & 4 s_3 \\
 0    &    0         &      0       &     s_4     & 4 s_2
\end{array}\right)\ .
\label{matrixMtwoa}
\ee
From this expression, it is easy to generalize to the general form for states
with $n$ original particles. Its eigenvalues and eigenstates, analogous to
\re{thebes}, \re{themus} can be read off explicitly in \re{orth2} and
\re{eigens}.

Note that, in the weak sense, there is a Hamiltonian $H^{[4]}$
which may be written directly in terms of the $d_j$ and $d_j^{\dag}$
\be
\left[H^{[4]},d_i^{\dag}\right]\mid 0>=M^{[4]}_{ji}d_j^{\dag}\mid 0>
\label{H4bsfour}
\ee
and is
\be
H^{[4]}=d_i^{\dag}N^{[4]}_{ij}d_j
\label{H4def}
\ee
with
\be
N^{[4]}=\frac{1}{6}\left( \begin{array}{ccccc}
 s_1  &  s_3        &0              &0            &0
			\\
 s_4  &3 s_1 +  s_2 &3 s_3          &0            &0
			\\
 0    &3 s_4        &3( s_1 +  s_2) &3 s_3        &0
			\\
 0    &0            &3 s_4          & s_1 + 3 s_2 & s_3
			\\
 0    &0            &0              & s_4         & s_2
\end{array}\right)\ .
\label{matrixNfour}
\ee

\item
Let us now try to define the $6\times 6$ matrix $\check{M}^{[4]}$ in the same
way for the $\check{d}_i^{\dag}$
\be
\left[H^{[2]}_{\rm{naive}},\check{d}_i^{\dag}\right]
       =\check{M}^{[4]}_{ji}\check{d}_j^{\dag}\ .
\label{Hbsfourdot}
\ee
again starting from the same canonical commutation relations for the
$b_i^{\dag}$'s and $\widetilde{b}_j$ \re{canonwtild}. We find
\be
\check{M}^{[4]}_{ij}=
\left[ \begin {array}{cccccc}
4s_1& s_3&0&0&0&0
\\
\noalign{\medskip}
4 s_4&3 s_1+s_2&2 s_3&2 s_3&0&0
\\
\noalign{\medskip}
0& s_4&2( s_1+ s_2)&0& s_3&0
\\
\noalign{\medskip}
0&2s_4&0&2( s_1+ s_2)&2s_3&0
\\
\noalign{\medskip}
0&0&2s_4&2s_4&s_1+3s_2&4s_3
\\
\noalign{\medskip}
0&0&0&0&s_4&4s_2
\end {array} \right] .
\label{Msix}
\ee

Since the $d_i^{\dag}$ are perfectly defined
there is no ambiguity. However, coherence with \re{twotwoid}
implies that an arbitrary combination of
$\check{d}^{\dag}_3- \check{d}^{\dag}_4$ can be added to the right-hand side of
\re{Hbsfourdot} and $\check{M}^{[4]'}_{ij}$ can be replaced by
\be
\check{M}^{[4]'}_{ij}=\check{M}^{[4]}_{ij}
            +r_{j}\delta_{i3}-r_{j}\delta_{i4}\ .
\label{MMprime}
\ee
namely
\bea
&&\check{M}^{[4]'}_{ij}=
    \nonumber\\
&&\hskip -5 mm
\left[ \begin {array}{cccccc}
4\, s_1& s_3& 0 & 0&0&0
                     \\
\noalign{\medskip}
4\,s_4 &3\,s_1+ s_2 &2\, s_3 &2\, s_3 &0 &0
                     \\
\noalign{\medskip}
r_1 & s_4+r_2 &2\,(s_1+ s_2)+ r_3 &r_4 &s_3+r_5 &r_6
                      \\
\noalign{\medskip}
-r_1 &2\,s_4-r_2 &- r_3 &2\,(s_1+ s_2)- r_4 &2\,s_3-r_5 &-r_6
                      \\
\noalign{\medskip}
0 &0 &2\,s_4 &2\,s_4 & s_1+3\,s_2 &4\,s_3
                       \\
\noalign{\medskip}
0 &0 &0 &0 & s_4 &4\,s_2
\end {array} \right] .
      \nonumber\\
&&
\label{sixxsix}
\eea

Making the same combination for the left-hand side,
it can easily be checked that
\be
\left[H^{[2]}_{\rm{naive}},\check{d}_3^{\dag}
-\check{d}_4^{\dag}\right]
=(2s_1+2s_2+r_3-r_4)(\check{d}^{\dag}_3- \check{d}^{\dag}_4)
\label{MMident}
\ee
i.e. a combination of $\check{d}^{\dag}_3- \check{d}^{\dag}_4$ only. This result is coherent when $\check{d}^{\dag}_3-\check{d}^{\dag}_4$  is put to zero.

The eigenvalues of the matrix $\check{M}^{[4]'}_{ij}$ depend upon
the values of the arbitrary parameters $r_i$ since the secular equation is
\be
(\lambda-E_{4,4})(\lambda-E_{4,-4})(\lambda-E_{4,2})(\lambda-E_{4,-2})
(\lambda-E_{4,0})(\lambda-E_{4,0}-r_3+r_4)
\label{eigenm4prime}
\ee
When $r_4 \neq r_3$, the eigenvalues are all different and the matrix can be
diagonalised. When $r_4=r_3$, two eigenvalues become equal.
Then there is always the eigenvector $(0,0,1,-1,0,0)^{tp}$. The condition for
the existence of a second eigenvector is
\bea
&&r_6  s_4^2 + 2 r_5 s_1 s_4 - 2 r_5 s_2 s_4 + r_3 (s_1 -s_2)^2 -  
2 r_3 s_3s_4 
    \nonumber\\
&&\phantom{xxxxxx}- 2 r_2 s_1 s_3 + 2 r_2 s_2 s_3 + r_1 s_3^2=0
\label{condtwoeigen}
\eea

When this condition is satisfied, the matrix can be transformed to a fully diagonal form; otherwise it is reducible to a Jordan Normal form

\be
\left[ \begin {array}{cccccc}
E_{4,4}& & 0 & 0&0&0
                     \\
\noalign{\medskip}
0&E_{4,2}&0 &0 &0 &0
                     \\
\noalign{\medskip}
0 & 0 &E_{4,0} &1 &0 &0
                      \\
\noalign{\medskip}
0 &0 &0 &E_{4,0}&0 &0
                      \\
\noalign{\medskip}
0 &0 &0 &0 & E_{4,-2}&0
                       \\
\noalign{\medskip}
0 &0 &0 &0 & 0&E_{4,-4}
\end {array} \right] .
\label{jordanf}
\ee

\end{itemize}

\section{Extension to a Fock space \label{secfock}}

We shall now promote the discussion to the case of a Fock-space Hamiltonian
linear in the products of one creation and one annihilation operator
$a^{\dag}_i a_j$ of bosons of two (or more) different species. The matrices $N$
now become functions of the creation and annihilation operators
$(a_j,\ a^\dag_j)$ which satisfy the usual canonical commutation relations
(see \re{com}).

In analogy with the 2-dimensional case \re{diaghamil} and with the discussion of
the preceding section, the diagonal Hamiltonian is defined as
\be
H_d=\sum_{j}\mu_j a_j^{\dag}a_j
\label{lmHdop}
\ee
and the related Hamiltonian $H$ by
\be
H=N H_d N^{-1}
\label{lmHop}
\ee
which should be linear in the products $a^{\dag}_i a_j$ since this
is a free field theory.
\noindent Suppose
\bea
N a_i N^{-1}&=&\sum_{j}c^{[1]}_{ij} a_j
       \quad\left(\equiv \alpha_i\right)
      \nonumber\\
N a_i^{\dag} N^{-1}&=&\sum_{j}d^{[1]}_{ij} a_j^{\dag}
       \quad\left(\equiv \alpha^{\dag}_i\right)\ .
\label{lmatf}
\eea

\noindent Indeed, we then find
\bea
H&=&\sum_{j}\mu_j
         \left(\sum_{k}d^{[1]}_{jk} a_k^{\dag}\right)
         \left( \sum_{m}c^{[1]}_{jm} a_m \right)
            \nonumber\\
 &=&\sum_{k}\sum_{m}\left(\sum_{j}\mu_j d^{[1]}_{jk} c^{[1]}_{jm}\right)
           a_k^{\dag}a_m\ .
\label{lmHinas}
\eea

\noindent If an interpretation in terms of creation and annihilation operators
is to
remain, the right hand sides of \re{lmatf} should obey the same commutation
relations \re{com}
as the $a$'s. This implies for the matrices $c$ and $d$ having respectively
$c_{ij}$ and $d_{ij}$ as
components the restrictions
\be
d^{[1]tp}=(c^{[1]})^{-1}\ .
\label{lmcdrelation}
\ee

\noindent After some algebra, using \re{lmcdrelation} and defining
\be
d^{[2]}=d^{[1]}-1\!\!1\ ,
\label{lmd2def}
\ee
one finds
\bea
\left[a_i,N\right]&=&d^{[2]tp}_{ij}\, N\,a_j
          \label{lmfinalNeq1}\\
\left[N,a_i^{\dag}\right]&=&d^{[2]}_{ij}\, a_j^{\dag}\, N \ .
\label{lmfinalNeq2}
\eea
Using a power expansion of $N$ in terms of the $a$'s, it is easy to show that
the quadratic part of $N$ is
\be
Q=\sum_{ij}d^{[2]}_{ij} a^{\dag}_j a_i\ ,
\label{lmQ}
\ee
and that the general $N$ is constructed from $Q$ as
\be
N=:\exp(Q):
\label{lmNexpQ}
\ee
where the symbol $:\ldots:$ denotes the normal product (annihilation operators
written at the right of creation operators). Indeed, we have
\bea
\left[a_i,N\right]&=&\partial_{_{a_i^{\dag}}} :\exp(Q):
              \nonumber\\
              &=&:\exp(Q):\left(\partial_{_{a_i^{\dag}}} Q\right)
              \nonumber\\
              &=&N\left(\sum_{j}d^{[2]}_{ji} a_j\right)
                   \quad\quad\Leftrightarrow
                   \quad\quad{\rm{Eq.}}\re{lmfinalNeq1}
\label{lnNproof1}
\eea
but also
\bea
\left[N,a^{\dag}_i\right]&=&:\partial_{_{a_i}} \exp(Q):
              \nonumber\\
              &=&:\left(\partial_{_{a_i}} Q\right)\exp(Q):
              \nonumber\\
              &=&\left(\sum_{j}d^{[2]}_{ij} a_j^{\dag}\right)N
                   \quad\quad\Leftrightarrow \
                   \quad\quad{\rm{Eq.}}\re{lmfinalNeq2}\ .
\label{lnNproof2}
\eea
This completes the proof.

It is then not difficult to see that, inversely that any $H$ of the form
\be
H=h_{ij}a_j^{\dag}a_i
\label{Hgeneric}
\ee
can be put into diagonal form $H_d$ by using the inverse of $N$ and adjusting
the
coefficient $c^{[1]},d^{[1]}$ suitably. The $\mu$'s are the eigenvalues of the
matrix constructed with the $h_{ij}$ which is supposed to be diagonalisable at
this stage.

In particular for the case of $p=2$, this can be
read off directly for an arbitrary $H$ from formulas \re{def4} \re{def5} \re{def6}
\re{def7} for the coefficient $c^{[1]},d^{[1]}$ of
Eq.\re{lmatf}.

Our arguments about the existence of a Hermitian Hamiltonian when the
eigenvalues are real also applies to the infinite dimensional case. While
we were concluding this article a paper by Mostafadazeh \cite{Mosta}
appeared, demonstrating that a quantum mechanical anharmonic oscillator with potential $ix^3$
giving rise to a non-Hermitian Hamiltonian can be transformed to a Hermitian
form, thus giving an alternative proof of the reality of energy
eigenvalues.

\section{Conclusions \label{secconclusion}}

The Landau problem of a particle in a magnetic field has been shown to
demonstrate the phenomenon of a non-Hermitian Hamiltonian giving rise to real
eigenvalues when the coupling to the magnetic field becomes pure imaginary has
been exhibited. All properties of this model are explicitly calculable as it is
really a free field theory in disguise, which illuminates the conditions under
which non-Hermitian Hamiltonians may have real eigenvalues. Indeed, starting
from real eigenvalues in a finite system, which is then subject to a similarity
transform, it has been shown how to define creation and annihilation operators,
and their conjugation to build a Fock space for a general non-Hermitian
Hamiltonian. In all the cases that we have studied in detail, we have observed
that, at least formally, when a non-Hermitian Hamiltonian possesses real
eigenvalues, there exists a change of basis such that the
transformed Hamiltonian in the corresponding subspace becomes
Hermitian Other properties of such systems have been discussed, notably the
conditions for degenerate eigenvalues. In the $2\times 2$ case and extensions
thereof this  is
a sign that the associated matrix is not fully diagonalisable
and marks the transition between real and complex conjugate pairs of
eigenvalues.

\section*{Acknowledgements}
One of us (D.B.F.) is indebted to Roman Jackiw and Patrick Dorey for discussions
on ${\cal{PT}}$ symmetry.

\end{document}